\documentclass[preprint,preprintnumbers,amsmath,amssymb]{revtex4}

\usepackage{graphicx}

\begin{document}
\title{Space Charge Effect and Mirror Charge Effect in Photoemission Spectroscopy}
\author{X. J. Zhou$^{1,2}$, B. Wannberg$^{3}$, W. L. Yang$^{1,2}$,
V. Brouet$^{1,2}$,  Z. Sun$^{4}$, J. F. Douglas$^{4}$, D.
Dessau$^{4}$, Z. Hussain$^{2}$ and Z.-X. Shen$^{1}$}

\affiliation{$^{1}$Dept. of Physics, Applied Physics and Stanford
Synchrotron Radiation Laboratory, Stanford University,  Stanford,
CA 94305
\\ $^{2}$Advanced Light Source, Lawrence Berkeley National
Lab, Berkeley, CA 94720
\\$^{3}$ Gammadata Scienta AB, P.O. Box 15120, SE-750 15 Uppsala, Sweden
\\ $^{4}$Department of Physics, University of Colorado, Boulder,
Colorado 80309-0390}

\date{\today}

\begin{abstract}

We report the observation and systematic investigation of the
space charge effect and mirror charge effect in photoemission
spectroscopy. When pulsed light is incident on a sample, the
photoemitted electrons experience energy redistribution after
escaping from the surface because of the Coulomb interaction
between them (space charge effect) and between photoemitted
electrons and the distribution of mirror charges in the sample
(mirror charge effect). These combined Coulomb interaction effects
give rise to an energy shift and a broadening which can be on the
order of 10 meV for a typical third-generation synchrotron light
source. This value is comparable to many fundamental physical
parameters actively studied by photoemission spectroscopy and
should be taken seriously in interpreting photoemission data and
in designing next generation experiments.

Key words:  Space charge, mirror charge, photoemission, Fermi
level shift, Fermi level broadening.

\end{abstract}


\maketitle


\section{Introduction}
Photoemission spectroscopy measures the energy distribution of
photo-emitted electrons when materials are irradiated with light
\cite{Huefner,SKeven}(Fig. 1). It is widely used in solid state
physics and chemistry for investigating the electronic structure
of surface, interface and bulk materials\cite{Huefner,SKeven}.
Recently it has become a prime choice of technique in studying
strongly correlated electron
systems\cite{ScienceIssue,SpecialJESRP}, such as high temperature
superconductors\cite{Damascelli}.  The availability of synchrotron
light sources and lasers, combined with the latest advancement of
electron energy analyzer, has made a dramatic improvement on the
energy resolution of photoemission technique in the last decade;
an energy resolution of $\sim$ 5meV or better can now be routinely
obtained. These achievements have made it possible to probe
intrinsic properties of materials and many-body
effects\cite{Damascelli}. For example, measurements of the
superconducting gap on the order of 1 meV, as in conventional
superconductors\cite{Chainani} and in some high temperature
superconductors\cite{Armitage}, have been demonstrated.


On the other hand, the utilization of pulsed light sources, such
as synchrotron light or pulsed lasers, has also brought about
concerns of the space charge effect\cite{Bjorn}. When a large
number of electrons are generated from a short pulsed source and
leave the sample surface, the electrons will first experience a
rapid spatial distribution depending on their kinetic energy.
Then, because of the Coulomb interaction, the fast electrons tend
to be pushed by the electrons behind them while the slow electrons
tend to be retarded by those fast electrons. This energy
redistribution will distort the intrinsic information contained in
the initial photoelectrons by giving rise to two kinds of effects.
One is a general broadening of the energy distribution, due to
both acceleration and retardation of electrons in their
encounters. The other is a systematic shift in the energy. The
space charge broadening of the energy distribution has been known
for a long time as a limiting factor in electron monochromators
and other electron beam devices\cite{Boersch}, but it has not been
considered in photoemission until very recently\cite{Bjorn}. The
main concern there was whether such an effect will set an ultimate
limit on further improving the energy resolution of the
photoemission technique\cite{Bjorn}.

Here we report the first experimental observation of the space
charge effect in photoemission. In addition, by combining
experimental measurement with numerical simulations,  we show that
the mirror charges (also known as image charges in the literature)
in the sample also play an important role in the energy shift and
broadening. The combined effect of these Coulomb interactions
gives an energy shift and broadening on the order of 10 meV for a
typical third-generation synchrotron light source, which is
already comparable or larger than the energy resolution set by the
light source and the electron analyzer. The value is also
comparable to the many-body effect actively pursued by modern
photoemission spectroscopy. These effects, therefore, should be
taken seriously in interpreting experimental data and in designing
next generation experiments.

\section{Experiment}
The experiment was carried out on beamline 10.0.1 at the Advanced
Light Source. This is a third-generation synchrotron source which
generates pulsed light with a frequency of 500 MHz and a duration
of $\sim$60 ps. The beamline can generate linearly-polarized
bright ultraviolet light with a photon flux on the order of
10$^{12}$ photons/second with a resolving power E/$\Delta$E of
10,000 (E is the photon energy and $\Delta$E the beamline energy
resolution). The endstation is equipped with a high resolution
Scienta 2002 analyzer. The analyzer, together with the chamber, is
rotatable with respect to the beam while the sample position is
fixed.  The measurement geometry is illustrated in the upright
inset of Fig. 1.  There are two angles to define the direction of
electrons entering the analyzer with respect to the sample normal:
tilt angle $\phi$ and analyzer rotation angle $\alpha$. We
measured the sample current to quantitatively measure the number
of electrons escaping from the sample which is proportional to the
photon flux.  With the pulse frequency of 500MHz at the ALS, 1 nA
of the sample current corresponds to 12.5 electrons per pulse.

Fig. 2a shows a typical photoemission spectrum of polycrystalline
gold taken with a photon energy of 35 eV. It consists of a Fermi
edge drop (E$_F$) near $\sim$30 eV, valence band between
20$\sim$30eV and a secondary electron tail extending to lower
kinetic energy arising from the inelastic scattering.  We chose to
measure on gold because the sharp Fermi edge at low temperature
($\sim$20 K for all the measurements in the paper) gives a good
measure of both the energy position and width (Fig. 2b). The Fermi
edge is fitted by the Fermi-Dirac function, {\it
f}(E)=1/$(exp(\frac{E-E_F}{k_{B}T})+1)$, at zero temperature
convoluted with a Gaussian with a Full-Width-at-Half-Maximum
(FWHM) $\Gamma$. This width $\Gamma$ includes all the
contributions from thermal broadening, analyzer resolution,
beamline resolution and others.

In photoemission experiments, it is a routine procedure to use
Fermi level of a metal (such as gold) as the energy referencing
point for the sample under study because the Fermi levels are
expected to line up with each other when the metal and the sample
are in good electrical contact. The Fermi level of the metal is
also expected to be dependent only on the photon energy and not on
other experimental conditions, such as sample temperature, photon
flux etc. It was therefore quite surprising when we first found
out that the gold Fermi edge shifts position with incident photon
intensity (Fig. 2b).  A systematic measurement reveals that, under
some measurement geometries, the Fermi level varies linearly with
the sample current and the shift can be as high as $\sim$20meV
within the photon flux range measured (Fig. 3a).  Note that the
Fermi level energy gets higher with increasing photon flux. This
rules out the possibility of sample charging that usually occurs
due to poor electrical grounding of the sample. In that case, the
Fermi level energy would be pushed downward with increasing photon
flux. We can also rule out the possibility of the local sample
heating due to high photon flux because temperature only affects
the Fermi edge broadening but will not change the Fermi level
position. As we estimated, for a photon flux of $\sim$10$^{13}$
photons/second at a photon energy of 35 eV, the corresponding
power is $\sim$0.056 mW.  The temperature increase with such a
small power, spread over an area of ~1 mm$^2$, is negligible so it
also has little effect on the thermal broadening of the Fermi
edge.

The first thing to check is whether this Fermi level shift with
photon flux is due to instrumental problems, which can be from
either the beamline or the electron analyzer. Regarding the
beamline, the photon flux is usually varied by adjusting the size
of the beamline slits. This will change the beamline energy
resolution correspondingly but may potentially also cause energy
position change. To check whether this is the case, we put a
photon blocker in the beamline (Fig. 1) so that it can attenuate
the photon flux while keeping the photon energy and resolution
intact. Using the photon blocker, we observed a similar variation
of the Fermi level with photon flux (Fig. 3a), thus ruling out the
possibility of beamline problems. We also put an electron blocker
(Fig. 1) to vary the number of electrons collected by the
analyzer. When the photon flux on the sample is fixed, the Fermi
edge shows little change with the number of electrons entering the
analyzer (Fig. 3b). This indicates that the energy shift we have
observed is not due to problems of the electron analyzer either.
Therefore, the observed energy shift must be associated with the
photoemission process itself.


In addition to the energy position shift, there is also an energy
broadening associated with increasing photon flux.  To observe
such an effect, we have to compromise the beamline energy
resolution in the way that it has a relatively high photon flux to
induce an obvious broadening effect,  and a relatively high energy
resolution ($\sim$10meV) in order to resolve the additional
broadening from all other contributions.  The measurement is made
possible by taking the advantage of the photon blocker to fix the
contribution from the beamline.  The total width increases with
increasing photon flux (inset of Fig. 4). Taking the width at the
lowest photon flux as arising from all the other contributions
including the beamline, the analyzer and sample temperature
broadening,  the photon-induced energy broadening can be extracted
after deconvolution. As seen in Fig. 4, it varies with the photon
flux with a magnitude comparable to but slightly larger than the
energy shift.


We have found that the Fermi edge shift and broadening are
sensitive to the spot size of the beam on the sample (Fig. 5).
Here the spot size is changed by varying the vertical focus of the
beamline; the horizontal beamsize is fixed. It is measured using
the transmission mode of the analyzer, calibrated by using samples
with known size.  As seen from Fig. 5a, as the spot size
increases, the energy shift gets less sensitive to the change of
photon flux, as also seen from the slope change as a function of
the spot size (Fig. 6).  For comparison, Fig. 6 also includes the
simulated data over a large range of spot sizes. Although the data
of energy broadening (Fig. 5b) is scattered as a result of
deconvolution from a relatively large background value, the trend
is clear that the broadening gets smaller with increasing spot
size. Again, for a given beam size, the magnitude of the energy
broadening is comparable to but slightly larger than the
corresponding energy shift.

The Fermi edge shift and broadening are also sensitive to the
electron emission angle.  We set the gold sample at different tilt
angles and measured the Fermi level position and width as a
function of the analyzer angle  under various photon flux.  As
seen in Fig. 7, the Fermi level position exhibits a strong
variation with the analyzer angle, particularly at high photon
flux. The Fermi level is higher near smaller analyzer angle and
decreases with increasing analyzer angle. When the analyzer angle
is close to 90 degrees all the curves with different sample tilt
angle and with different sample current tend to approach to a
similar position within the experimental error. The overall
measured Fermi level width basically follows the trend of the
energy shift: it becomes smaller with increasing analyzer angle.
We also notice that the curves are not symmetrical with respect to
the zero analyzer angle. Since the surface of the polycrystalline
gold we used is not perfectly flat, one possible reason is that
the exact angle may be slightly off from the nominal value.
Another possibility is the presence of a small systematic error.
As indicated from Fig. 7, when the sample current is small (23
nA), one can still observe Fermi level shift with the analyzer
angle which may be due to a systematic error associated with the
experimental setup.

To gain more insight on the angle dependence, we also measured the
energy shift and broadening as a function of the sample current at
different analyzer angles (Figs. 8a and b).  It is interesting to
note that, while for small analyzer angles, the energy shift is
proportional to the sample current, as we have seen before, it
deviates significantly from the straight line for large angles. In
this case, the energy shift exhibits linear relation only at high
sample current. When the sample current gets smaller, it goes
through a minimum, and then gets larger again even with further
decreasing of the sample current. One may expect that at zero
sample current the energy shift approach zero so that all curves
should converge at the zero sample current, as indeed shown by the
data in Fig. 8 (the small Fermi level scattering at zero sample
current may be due to the systematic error as discussed before).
This implies that, for large analyzer angles, the energy shift can
be even negative at some sample current.

Fitting the high sample current part of the curves in Figs. 8a and
b with a straight line, we extracted their slopes and plotted them
in Fig. 8c for two sample tilt angles. The shape of the curves is
similar to that in Fig. 7. The high sample current part overlaps
with each other. When extrapolated to 90 degrees the Fermi level
shift is approaching zero which is also consistent with the
converging of the Fermi level at high analyzer angle as seen in
Fig. 7.

To further investigate the origin of the angle-dependent energy
shift and broadening, we measured the gold valence band at
different analyzer angles (Fig. 9). The intensity of these spectra
are normalized to the photon flux so they are comparable with each
other. The shape of the valence band shows no obvious change with
the analyzer angle, but their relative intensity changes
dramatically. For a quantitative comparison, we integrated the
spectral weight over a large energy range (5$\sim$35 eV) and the
result is shown in the inset of Fig. 9. Integration over a smaller
energy window such as 25$\sim$35 gives essentially the same shape.
We have found that the angular variation of the relative valence
band intensity and the Fermi level shift is identical (inset of
Fig. 9). This indicates that the angle dependence of the Fermi
level is directly related to the angle-dependence of the number of
photo-emitted electrons.

\section{Numerical simulation of space charge effect and mirror
charge effect}

It is expected that the space charge effect depends on a number of
parameters\cite{Bjorn}: (1). the number of electrons per pulse;
(2). the pulse length; (3). the size and shape of the excitation
area; and (4). the energy distribution of the electrons.  We have
performed numerical simulations using the Monte Carlo-based
technique developed earlier\cite{Bjorn} in order to quantitatively
examine our results. This serves first to check whether the
observed energy shift and broadening can be entirely attributed to
the space charge effect. It then helps to understand the
microscopic processes associated with it, such as the time scale
of the process. Moreover, it can be extended to investigate
situations that are difficult or not accessible for the
experiments, such as the effect of the electron energy
distribution, the effect of the pulse length,  and the case of a
continuous source, as we will discuss below.

In the simulation, a specified number of electrons ( 1-100000)
(denoted as interaction electrons hereafter) are started at random
positions within the specified source area, at random times during
the pulse, and with random energies with some specified
distribution. Because the acceptance angle of the electron energy
analyzer is small, the electrons for which the energy spread and
broadening are to be calculated (denoted as test electrons
hereafter) are started in the forward direction with a specified
initial energy but with a random distribution in start position
and time. This condition corresponds to the measurement geometry
of the analyzer angle $\alpha$=0 and the sample tilt angle
$\phi$=0. Each test electron is assumed to feel the Coulomb force
from all interaction electrons within some cut-off distance. The
interaction electrons are assumed to move in straight lines
defined by their initial conditions, i.e. all mutual interactions
between them are neglected. This is legitimate because their
position changes are extremely small and random. The energy
evolution of a single test electron is followed until all
interaction electrons have vanished outside the cut-off distance.
Then, the process is repeated with a new set of interaction
electrons and one new test electron. This procedure is repeated a
few thousand times, after which the energy distribution of the
test electrons is calculated. For the accuracy of the integration
to be of the same order of magnitude as the statistical
uncertainty, the cut-off distance has to be at least 1 mm, and for
most calculations it was chosen to be 2 mm. The energy
distribution can usually be well fitted by a Gaussian, although
the number of electrons which experience very large shifts is
significantly larger than for the Gaussian distribution. Such
extreme outliers are neglected when calculating the width of the
distribution.

The electrons in the pulse will experience Coulomb interaction
from all the other electrons at different energies, including the
large number of low-energy secondary electrons (Fig. 2a). To
evaluate the effect of the electron energy distribution on the
electrons at the Fermi level , we divided the energy range below
E$_F$ into a number of regions, and calculated the contribution
from each individual region. The simulated energy shift and
broadening from the direct space charge effect are plotted in
Figs. 10a and 10b, respectively. The energy shift displays a
strictly linear relation with the number of electrons in  a pulse
and the slope as a function of test electron kinetic energy is
plotted in Fig. 11. On the other hand, the energy broadening
exhibits a nearly linear relation only at large number of
electrons; at small number of electrons it shows a bend. Clearly
all electrons contribute to the Fermi level energy shift and
broadening but they contribute differently: the high-energy
electrons contribute more than the low-energy ones (Fig. 11).

In fact, an electron at a distance {\it z} in front of a
conducting metal surface will also experience an attractive force
F(z)=-e$^2$$/$(2z)$^2$, identical to that produced by a positive
(mirror image) charge at a distance {\it z} inside the
metal\cite{UHofer}. The basic assumption behind the mirror charge
concept is that the charges on the sample surface redistribute
themselves in such a way that the surface is always an
equipotential surface. Whether this assumption is correct on the
time scale considered here may be dependent,  e.g.,  on the
conductivity of the sample. In this case, each interaction
electron is accompanied by a mirror charge in the sample (inset of
Fig. 12a), which also interacts with the test electron. The
interaction of the test electron with its own mirror charge is not
included here because it is always present. In the earlier
simulation\cite{Bjorn}, the mirror charges could be neglected when
only considering the broadening caused by interaction electrons
with energies close to that of the test electron. For the case
when the test electron has higher energy than all interaction
electrons, this is no longer true, in particular when the energy
shifts are also considered.

Fig. 12a and 12b show simulated energy shift and broadening for
different energy ranges by incorporating both the space and mirror
charge effects.  The energy shift retains a linear variation with
the number of test electrons per pulse and the slope is plotted in
Fig. 11. The contribution from the mirror charge alone can be
easily extracted. Apparently the mirror charge gives rise to a
negative energy shift with increasing number of electrons per
pulse. This helps in compensating the positive energy shift from
the space charge effect. The combined effect on the energy
broadening is more complicated. For the highest energy range of
the interaction electrons (25 - 30 eV), the combined broadening
(Fig. 12b) is larger than that from the space charge effect alone
(Fig. 10b). But for the lower energy range of the interaction
electrons, it is smaller than that from the space charge effect.


We have found that the energy shift and broadening occur at very
different time scales.  As seen from Fig. 13, the energy shift
evolves gradually within the first nanosecond. The energy
broadening, on the other hand, has already reached its equilibrium
value at 100 ps, followed by random fluctuations. This is because
the energy shift takes place only after the electrons have
spatially sorted themselves according to their energy; after that
the forces are all acting in the same direction. We also note that
initially each interaction electron and its mirror charge form a
very short dipole, from which the field decreases rapidly with
distance. The broadening, on the other hand, is much more of a
nearest-neighbor effect, which is strongest when the pulse is
dense. Detailed study of the energy evolution for individual
electrons shows that the random part of the energy change is often
dominated by one single event, i.e., a close encounter with
another electron. Since the energy shift continues to grow over a
time that is comparable to the interval between pulses, we have
also checked whether it can be affected by remaining slow
electrons from the previous pulse: we have found that this
contribution is completely negligible.

Since a time-continuous light source, such as discharge lamps, is
widely used as a lab source for photoemission, it is important to
check whether similar effects still exist in that case.  For a
continuous light source, because there will be no spatial
redistribution of the electrons according to their energy, one
might expect the contribution to the energy shift from the space
charge  to be close to zero, while the mirror charge will give a
negative shift. The broadening can be expected to be of the same
order of magnitude as that from a pulsed source with the same
number of electrons per unit time. To simulate a continuous
source, we first start with a pulsed source, varying the pulse
length while keeping the number of electrons per unit time
constant, and try to extrapolate to infinite length to approximate
a continuous source. We have considered a typical case of Helium I
radiation (photon energy 21.12eV) on polycrystalline gold, and
varied the sample current during the pulse from 0.15 to 50
electrons/ps. Fig. 14 shows the energy shift and broadening for
different sample currents as a function of the pulse length. When
scaled by the sample current, all energy shift curves overlap with
each other because the shift is proportional to the current for
all pulse lengths (Fig. 14a). The energy shift shows non-monotonic
dependence with the pulse length, owing to the competition between
the direct space charges and mirror charges. When the pulse length
is short, the space charge dominates which gives positive energy
shift. When the pulse length is long enough, the effect from
mirror charges dominates which leads to a negative energy shift.
Eventually it asymptotically reaches a value that can be taken for
a continuous source. The shift is -0.7meV for 1.5$\times10$$^{12}$
electrons/second and can get significant when the photon flux is
larger. The energy broadening (Fig. 14b), on the other hand, does
not scale with the sample current, particularly at longer pulse
length. It also exhibits a non-monotonic variation with the pulse
length, reaching a maximum around 10$^{4}$ ps and then decreases
with further increasing of the pulse length.  If we assume an
asymptotic behavior following the drop, the broadening for the
continuous source is close to 1 meV for a sample current of
1.5$\times$10$^{12}$ electrons/second.

\section{Comparison between the experiment and the numerical simulation and discussions}

As we have seen from both the experimental measurements and the
simulation, the energy shift and broadening depend on many
parameters, such as the number of electrons per pulse, the pulse
length, the spot size on the sample, the emission angle of
electrons and the photon energy used. Moreover, it is
material-specific. This is first because it depends on the shape
of the valence band, i.e, the energy distribution of
photoelectrons. Seocnd, for metals and insulators, the effect of
mirror charge may vary significantly. With so many factors coming
into play simultaneously, it is hard to exhaust all the
possibilities and a proper approach to take is to measure or
simulate on an individual basis.

As shown in Figs. 4 and 5, the measured energy shift is
proportional to the sample current and the broadening is nearly
linear at high sample current and shows a bend at lower sample
current. Qualitatively speaking, both observations are consistent
with the simulated results from either the space charge effect
(Fig. 10) or combined space and mirror charge effects (Fig. 12).
After obtaining the contribution for each individual energy range
from the space charge effect (Fig. 10), we calculated the overall
energy shift from the measured valence band (Fig. 2a) as a
weighted sum of the contributions from the different energy
ranges. We also used a model where the energy distribution is
approximated by a rectangular shape corresponding to the valence
band and a triangular distribution of the secondary electrons; the
obtained results are similar. It was found that the value for the
energy shift obtained from the space charge effect alone is much
higher than that measured from experiment. For example, for the
spot size of 0.43mm$\times$0.42mm, the calculated energy shift is
0.175 meV/nA, much higher than the measured 0.055 meV/nA.  The
large discrepancy indicates that the space charge effect alone can
not account for the observed energy shift. This prompted us in
identifying the mirror charge effect that should be present for
metals such as gold. After considering both effects(Fig. 12), the
calculated energy shift becomes quantitatively consistent with the
experiment, as seen in Fig. 6 even for different spot sizes.
Considering that there are no adjustable parameters in the
simulation, this level of agreement is striking. This indicates
that we have captured the main contributors to the energy shift
effect. The quantitative comparison between the measurement and
the simulation has made us able to identify the mirror charge
effect that was not included before\cite{Bjorn}.

For the area dependence (Fig. 6), we note that the size of the
spot on the sample relative to the distance an electron travels
during the pulse is important. Depending on the relative ratio,
the space charge effect may exhibit different dependence on the
spot size. If the light spot is much larger than the electron
travelling distance (for 30 eV electron, the travelling distance
is $\sim$0.2 mm within 60 ps), the shape of electron spatial
distribution is basically flat. The space charge effect is
expected to be proportional to the number of electrons/area. When
the  spot size gets smaller, one will get increasingly important
edge effects, because electrons that move outside the spot will
not be compensated by electrons coming from the outside. In the
limit where the spot is very small, the spatial distribution of
electrons is a half-sphere. The average distance between electrons
will be defined by their time interval rather than by the distance
between the points where they started. So in that case, the effect
may become independent of the spot size.

On the other hand, there are cases where the simulation deviates
from measurements.  We found that the measured broadening is
larger than the values calculated from the simulation. As shown in
Fig. 6, from the simulation, the broadening is smaller than the
shift whereas from the measurement (Fig. 4 and 5), the broadening
is comparable or slightly larger than the shift. The reason for
this discrepancy is not clear yet and probably more sophisticated
simulations are needed to address the discrepancy.  We note the
broadening can be larger than the shift when the energy of the
interaction electrons is close to that of the test electron
(energy range 25$\sim$30 eV in Fig. 12b) which is probably due to
the longer average interaction times.  However, in the case of
gold, because the fraction of electrons in the range close to the
Fermi edge is very small (Fig. 2a), this contribution is small to
the overall broadening.   The angular dependence of the energy
shift (Fig. 7) can be well attributed to angle-dependent number of
electrons at different emission angles (Fig. 9) which is probably
associated with the linear polarization of the synchrotron light.
However, to understand the negative energy shift for high analyzer
angles at lower sample current, more simulation is also needed.

\section{Implications of space charge effect}


The observation of space and mirror charge effects has important
implications in photoemission experiments as well as the future
development of the technique.  These findings first ask for
particular caution in interpreting photoemission data. One
immediate issue is the electron energy referencing in
photoemission spectroscopy.  In photoemission community it is a
routine procedure to use the Fermi level of a metal as a
reference. This is usually realized by measuring the Fermi level
from a metal (such as gold) which is electrically connected to the
sample under measurement. It is true that the intrinsic Fermi
level of the sample is lined up with that of the metal, but the
measured Fermi level has an offset from the space and mirror
charge effects. This offset can be different between the sample
and the metal because it is not only material-specific, but also
depends on many other factors. When the effect on the energy shift
is strong, using the Fermi level from a metal as a reference
becomes unreliable.

Another related issue is the Fermi level instability during
measurement. Because the photon flux usually changes with time for
many synchrotron light sources due to the finite life-time of
electrons in the storage ring, the Fermi level is always changing
with time during measurements. As we have shown before, this can
give rise to an Fermi level uncertainty on the order of 10 meV for
a typical experimental setting using a third-generation
synchrotron light source. This is comparable or larger than many
energy scales which are actively pursued in many-body problems in
the condensed matter physics\cite{ScienceIssue,SpecialJESRP}.
Measurement with an energy precision of 1 meV is necessary, for
example, when the superconducting gap in some conventional metals
as well as in some high temperature superconductors is on the
order of 1 meV\cite{Chainani,Armitage}. In this case, an
uncertainty or shift on the order of 10 meV definitely poses a big
problem.

To resolve the Fermi level referencing problem, one can always
minimize the space charge effect by reducing the photon flux, or
increasing the spot size.   Apparently this is not desirable,
particularly when a high photon flux is necessary to take data
with a good statistics and a high efficiency. Given that the Fermi
level referencing to a metal is no longer reliable,  one may use
an internal reference from the sample under measurement. This
internal reference can be obtained from {\it priori} knowledge or
measurements with negligible space charge effect. For example, in
high temperature cuprate materials, the (0,0) to ($\pi$,$\pi$)
nodal direction can be used as an internal reference to locate the
Fermi level because it has been shown that the superconducting gap
and pseudogap approaches zero along this direction except for
slightly doped samples\cite{Damascelli}. As for the Fermi level
instability with time, since the energy shift exhibits a linear
relation with the photon flux, it is possible to make corrections
by recording the sample current or photon flux. Ideally, this
problem can be minimized if the synchrotron light source is
operated at a constant or quasi-constant photon flux (``Top-off")
mode.

In addition to the Fermi level uncertainty, the energy broadening
is another serious issue facing the photoemission technique. Since
most physical properties of materials are dictated by electronic
excitations within an energy range of $\sim$k$_B$T near the Fermi
level (k$_B$ is the Boltzman constant and T a temperature),  to
probe the intrinsic electronic properties, the energy resolution
has to be comparable or better than k$_B$T, which is 0.8 meV for
10 K.  Therefore, there is a strong scientific impetus to improve
the photoemission technique to even higher energy resolution
(sub-meV), accompanied by high photon flux and small beam size.
The space and mirror charge effects should be taken into account
seriously in the future development of new light sources and
electron energy analyzers.  The high photon flux and small spot
size will enhance the space and mirror charge effects; the
resultant energy broadening can be well beyond the resolution from
the electron analyzer and the light source.

With the increasing demand of high energy resolution, it is
important to investigate how to alleviate or remove the space
charge effect. For example, it is interesting to study whether
applying a bias voltage between the sample and the electron
detector will affect the space charge effect. On the other hand,
in addition to putting more effort on improving the performance of
the light sources, it is very important to put emphasis on
enhancing the capabilities of the electron energy analyzer. One
aspect is to further increase the sensitivity of electron
detection by using new electron detection schemes. The other
aspect is to keep improving the analyzer throughput. Note that
even for the state-of-the-art display electron analyzer, using
angle-resolved mode, only less than 1$\%$ of electrons are
collected during measurements while all the rest of electrons
emitted over 2$\pi$ solid angle from the sample surface are
wasted. A new scheme needs to be explored on how to record large
solid angle at the same time when maintaining high energy
resolution.  It is apparent that much work needs to be done and we
hope our identification of the Coulomb effects can stimulate more
work along this direction.

\newpage

\begin{figure}[tbp]
\begin{center}
\includegraphics[width=0.90\columnwidth,angle=0]{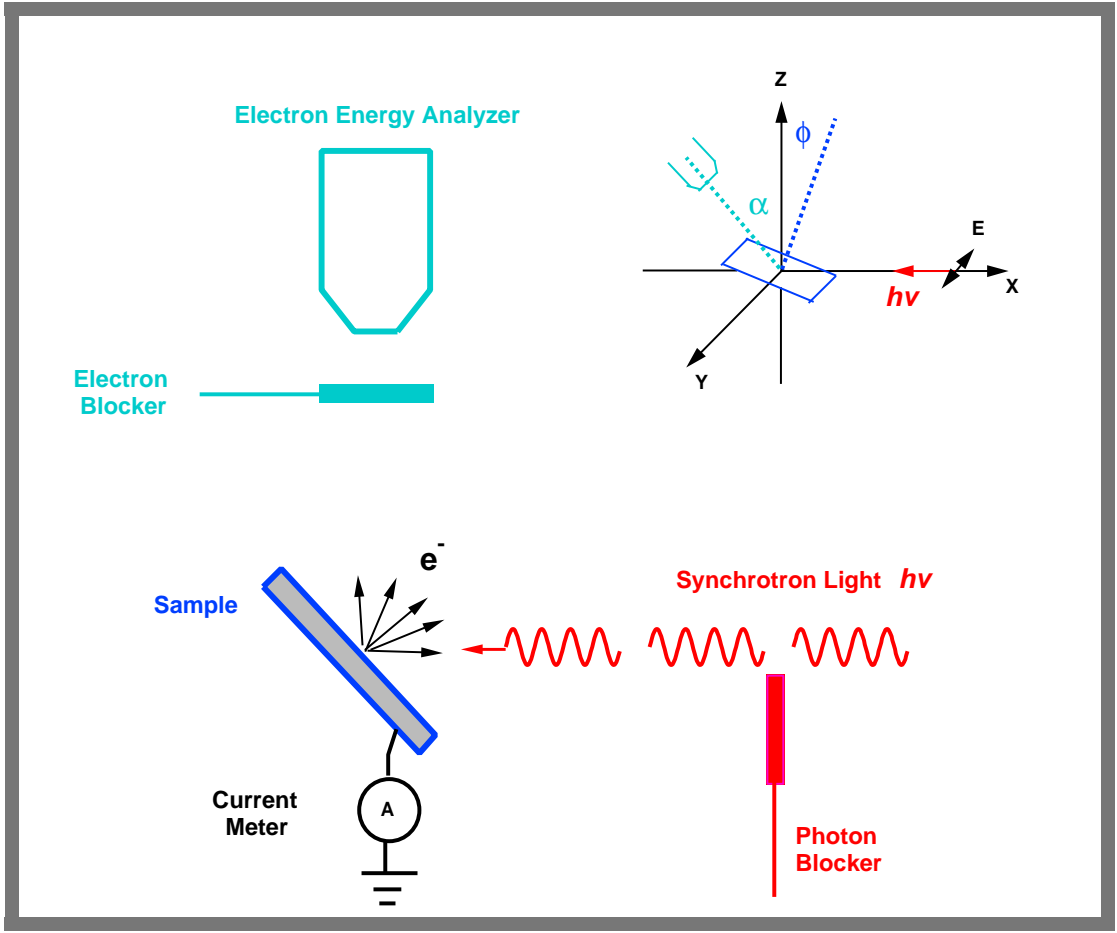}
\end{center}
\caption{Schematic of photoemission setup. A pulsed light is
incident on the sample, kicking out electrons, and the electrons
are collected by the electron energy analyzer. The photon blocker
is used to change photon flux while keeping the beamline intact.
The electron blocker is used to change the number of electrons
collected by the analyzer. The sample current recorded by a
picoammeter measures the number of electrons out of the sample
which is proportional to the photon flux. In the upright inset
shows the measurement geometry of the light, the sample and the
analyzer. The synchrotron light is along the X axis, with its
electrical field $\vec{E}$ in the XY horizontal plane and parallel
to Y axis. The sample normal is in the XZ plane and its angle with
respect to the Z axis is referred to as $\phi$. The analyzer is
rotatable and the lens axis is in the YZ plane. The angle of the
lens axis with respect to the Z axis is referred to as $\alpha$.}

\end{figure}

\begin{figure}[tbp]
\begin{center}
\includegraphics[width=0.65\columnwidth]{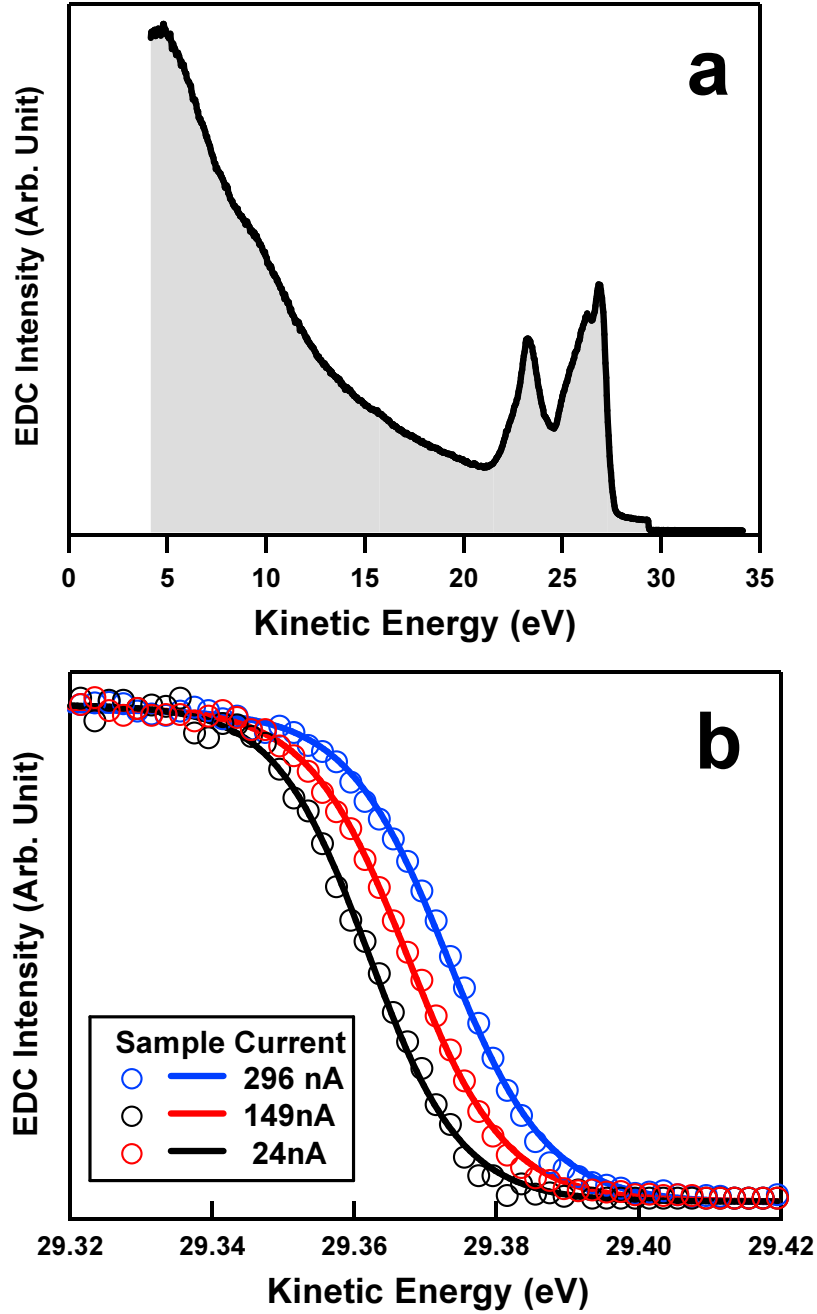}
\end{center}
\caption{Photoemission spectra of a polycrystalline gold measured
at a photon energy of 34 eV and a temperature of 20K. (a). Large
energy range spectrum showing a Fermi cutoff at 29.38eV, the
valence band between 20 and 30 eV, and lower energy part arising
from secondary electrons. (b). Au Fermi level measured at
different photon flux, as indicated by different sample current.
The open circles are experimental data which are fitted by
Fermi-Dirac functions (lines).
 }
\end{figure}

\begin{figure}[tbp]
\begin{center}
\includegraphics[width=0.85\columnwidth,angle=-90]{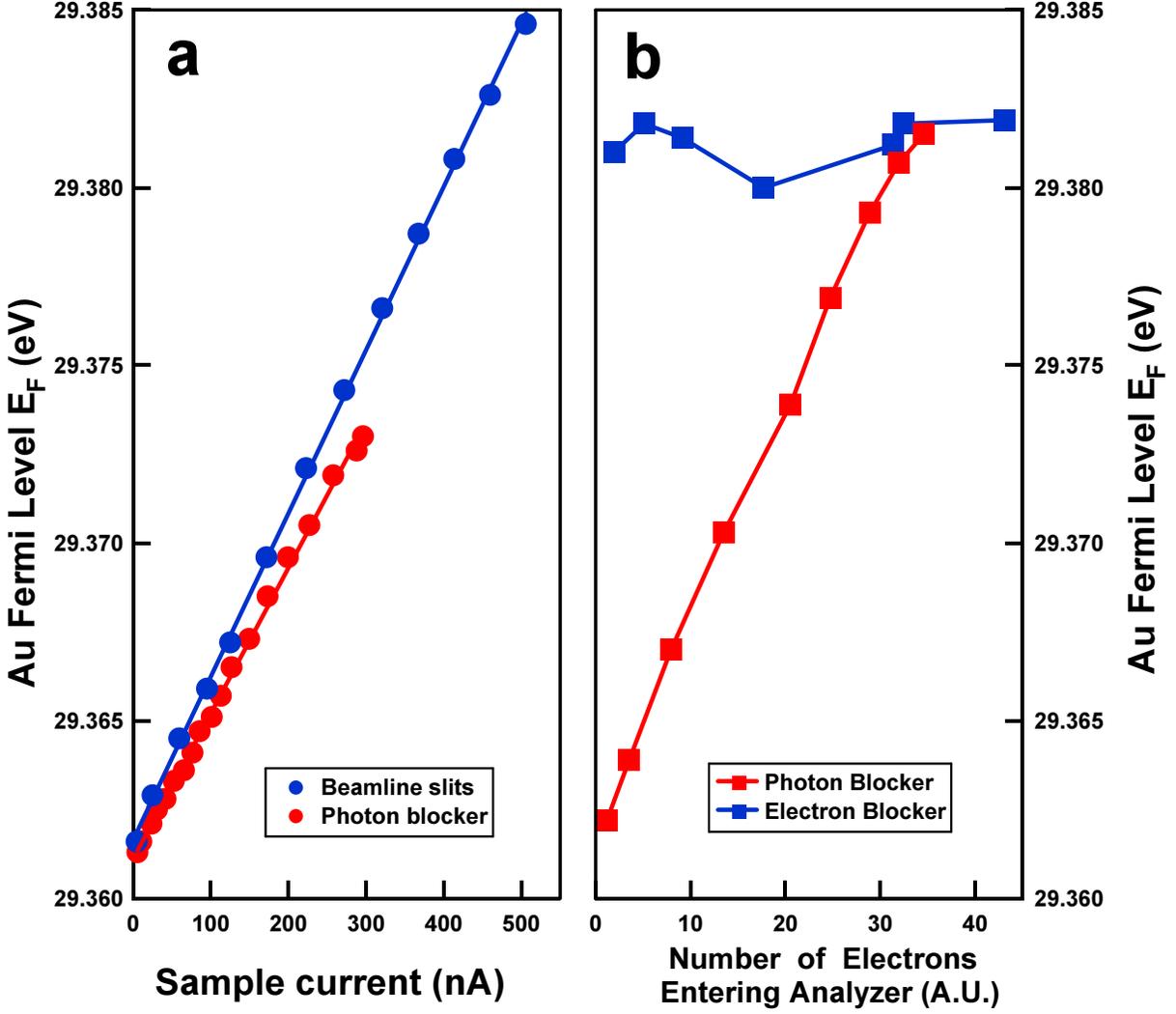}
\end{center}
\caption{Fermi level shift in photoemission process. The sample
tilt angle $\phi$ is 45 degrees and the analyzer angle $\alpha$ is
0. (a). Fermi level shift with the sample current varied by either
varying the beamline slits (blue circle) or by using the photon
blocker (red circle). (b). Fermi edge shift as a function of the
number of electrons entering the analyzer varied  by either
changing the photon flux by using the photon blocker(red square)
or keeping the photon flux constant but using the electron blocker
(blue square).
 }
\end{figure}

\begin{figure}[tbp]
\begin{center}
\includegraphics[width=0.8\columnwidth]{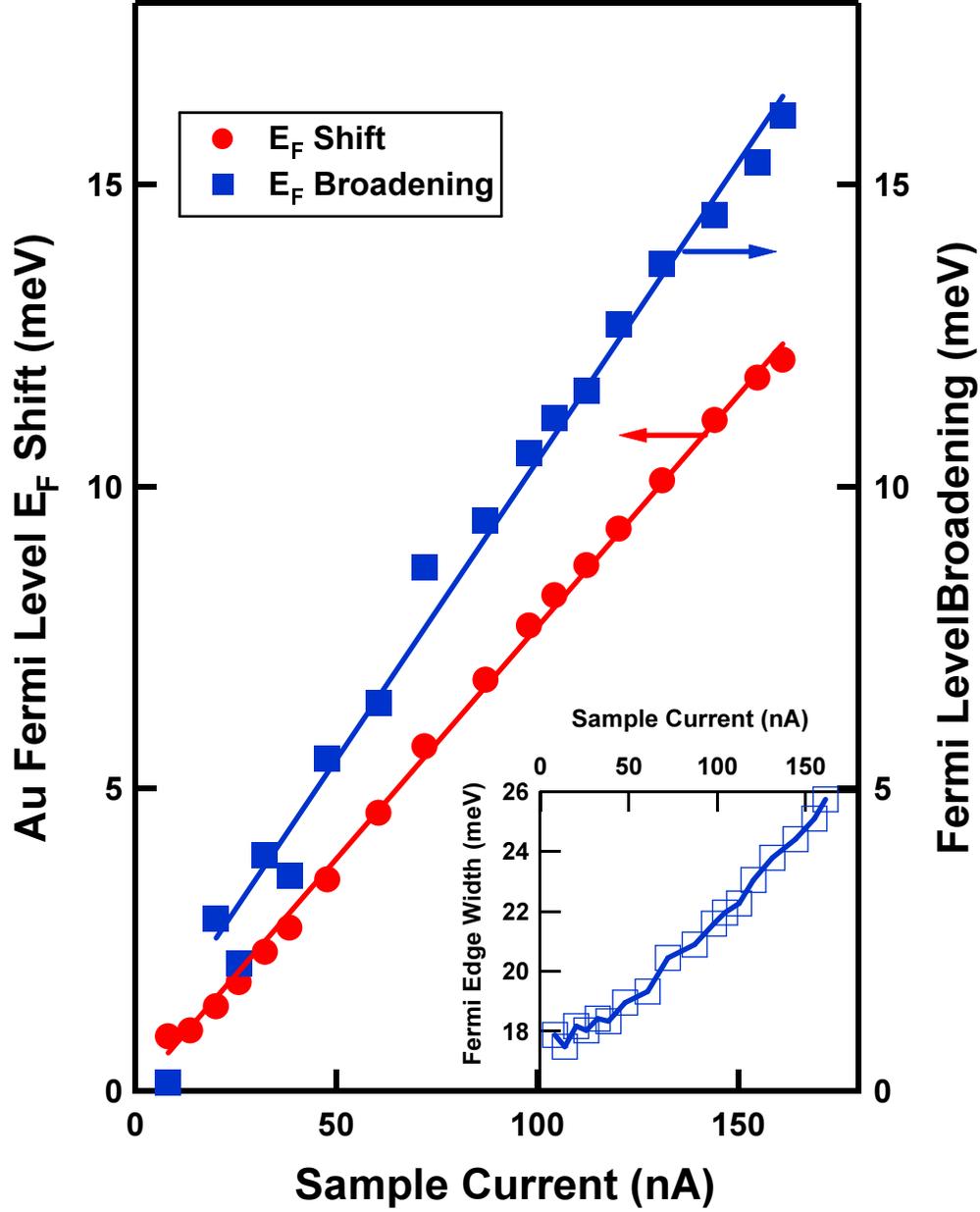}
\end{center}
\caption{Fermi edge broadening (blue square) and the Fermi edge
shift (red circle) as a function of sample current. The sample
tilt angle $\phi$ is 45 degrees and the analyzer angle $\alpha$ is
0. The beam spot size is $\sim$ 0.43$\times$0.30 mm$^2$. The
photon flux corresponding to 150 nA sample current is
$\sim$5$\times$10$^{13}$ photons/second.  The inset shows the
measured overall Fermi edge width as a function of the sample
current,  which includes all contributions including the beamline,
the analyzer and the temperature broadening. The net broadening
resulting from pulsed photons is obtained by deconvolution of the
measured data, taking the width at low photon flux as from all the
other contributions.
 }
\end{figure}

\begin{figure}[tbp]
\begin{center}
\includegraphics[width=0.60\columnwidth]{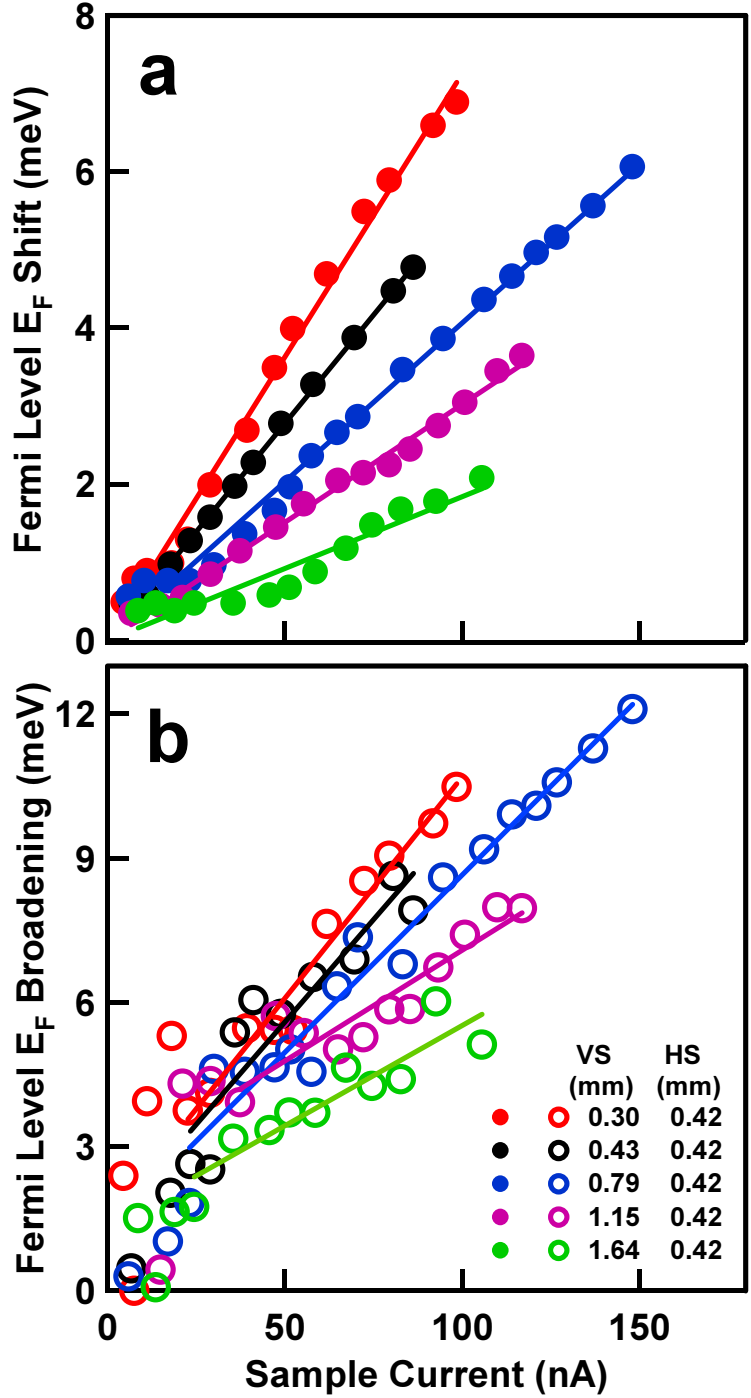}
\end{center}
\caption{(a). Effect of beam spot size on the energy shift. The
energy shift for each spot size (FWHM) shows a nearly linear
dependence on the sample current and  is fitted with a straight
line (solid lines). The sample tilt angle $\phi$ is 45 degrees and
the analyzer angle $\alpha$ is 0.  (b). Effect of beam size on the
energy broadening. The broadening at high sample current can be
approximated as a straight line; the solid lines also act as a
guide to the eye.}
\end{figure}

\begin{figure}[tbp]
\begin{center}
\includegraphics[width=0.8\columnwidth]{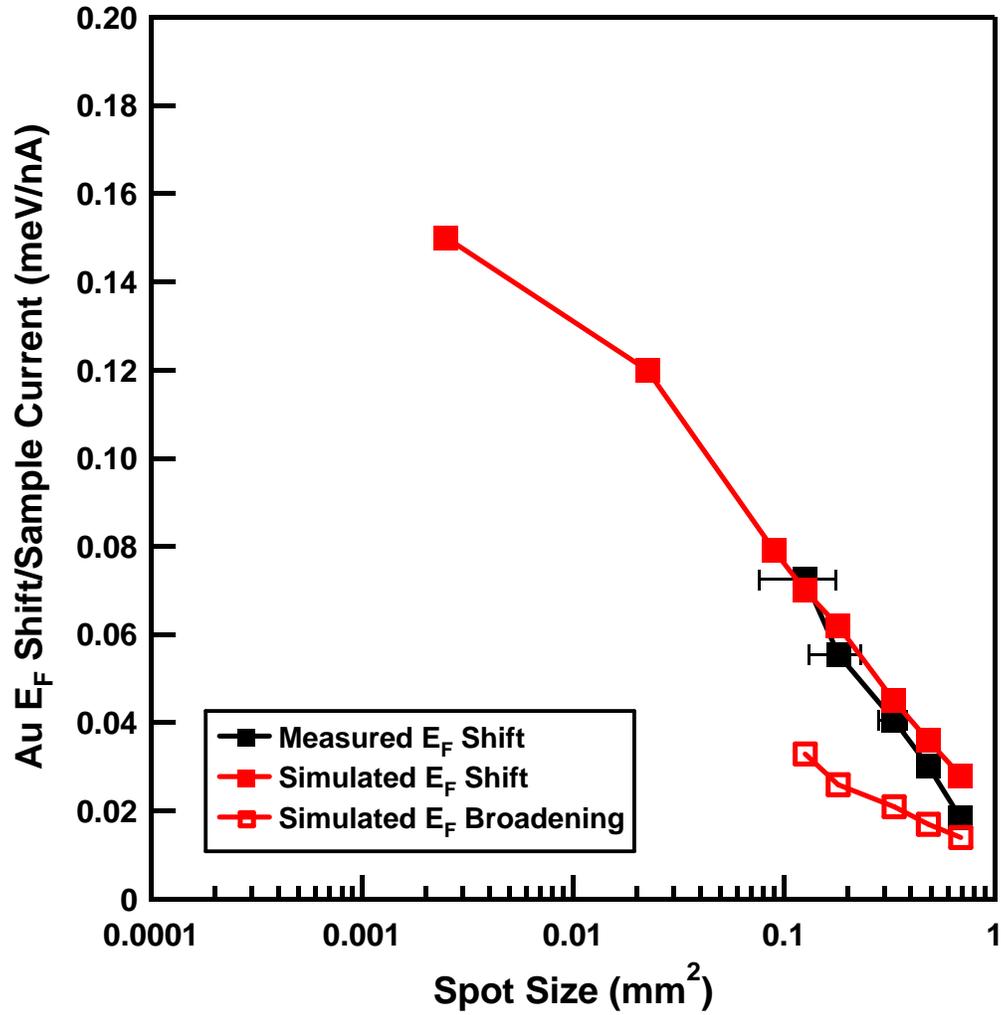}
\end{center}
\caption{Effect of beam spot size on the energy shift and
broadening. The measured energy shift (black solid square) is
obtained from the slope by fitting the curves in Fig. 5a with
linear lines. The simulated energy shift (red solid square) and
broadening (red open square), are calculated by considering both
the space and mirror charge effects and the overall gold valence
band as shown in Fig. 2a. }
\end{figure}

\begin{figure}[tbp]
\begin{center}
\includegraphics[width=0.92\columnwidth]{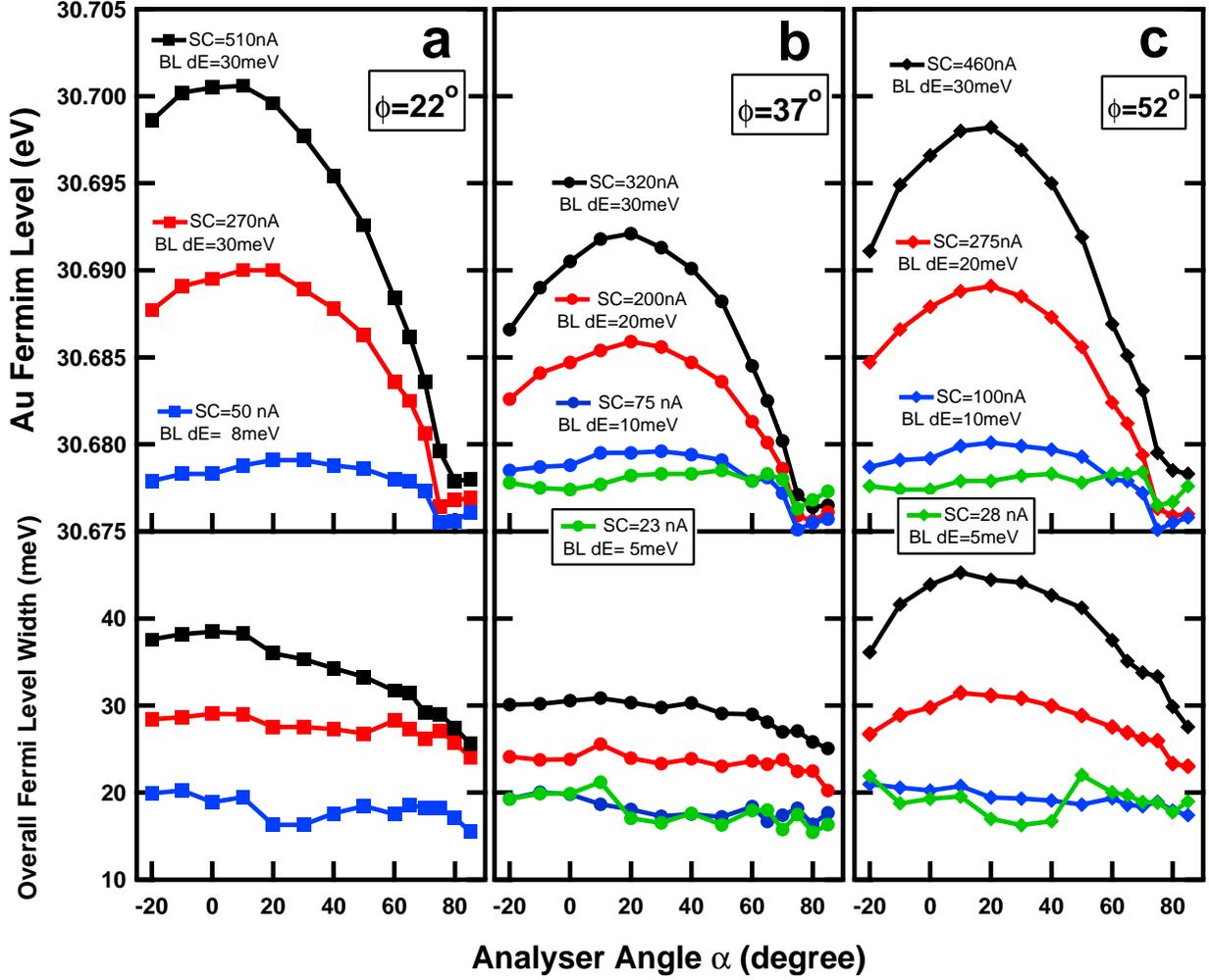}
\end{center}
\caption{The energy shift (upper panel) and broadening (lower
panel) as a function of the analyzer angle $\alpha$ for different
sample tilt angles (a).$\phi$=22 , (b).$\phi$=37  and
(c).$\phi$=52 . The curves in each panel represent different
sample currents (SC) under a given beamline resolution (dE).  For
any given curve the sample current is nearly a constant. }
\end{figure}

\begin{figure}[tbp]
\begin{center}
\includegraphics[width=0.56\columnwidth,angle=-90]{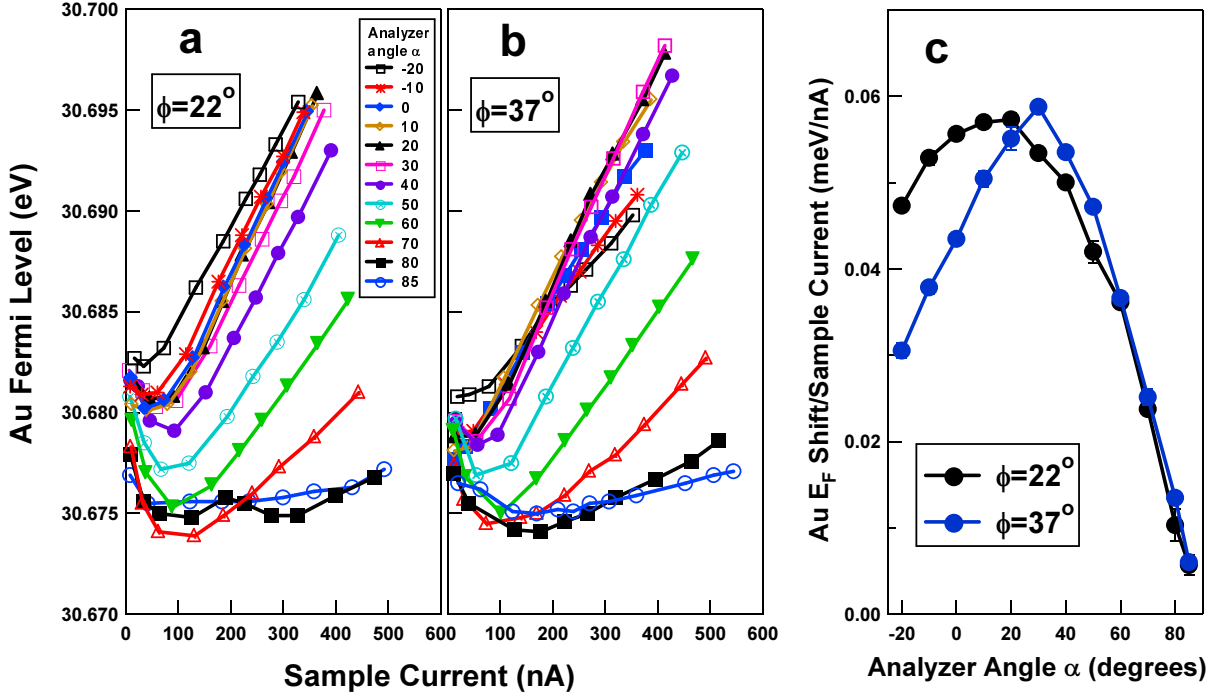}
\end{center}
\caption{Energy shift as a function of sample current at different
analyzer angles for the sample tilt angle of (a).$\phi$=22 degrees
and (b).$\phi$=37 degrees. The spot size for these two cases is
approximately 0.4mm$\times$0.4mm.  Note that while for the small
analyzer angle the energy shift changes nearly linearly over the
entire sample current range, for large analyzer angle, it shows a
back bend at low sample current. If we take the value near zero
sample current as the intrinsic Fermi level, it is found that the
energy shift at high analyzer angle can be negative. (c). Energy
shift per sample current for two different sample tilts angles.
The slope is obtained by fitting the straight high sample current
part as in Figs. 8a and 8b.
 }
\end{figure}

\begin{figure}[tbp]
\begin{center}
\includegraphics[width=\columnwidth]{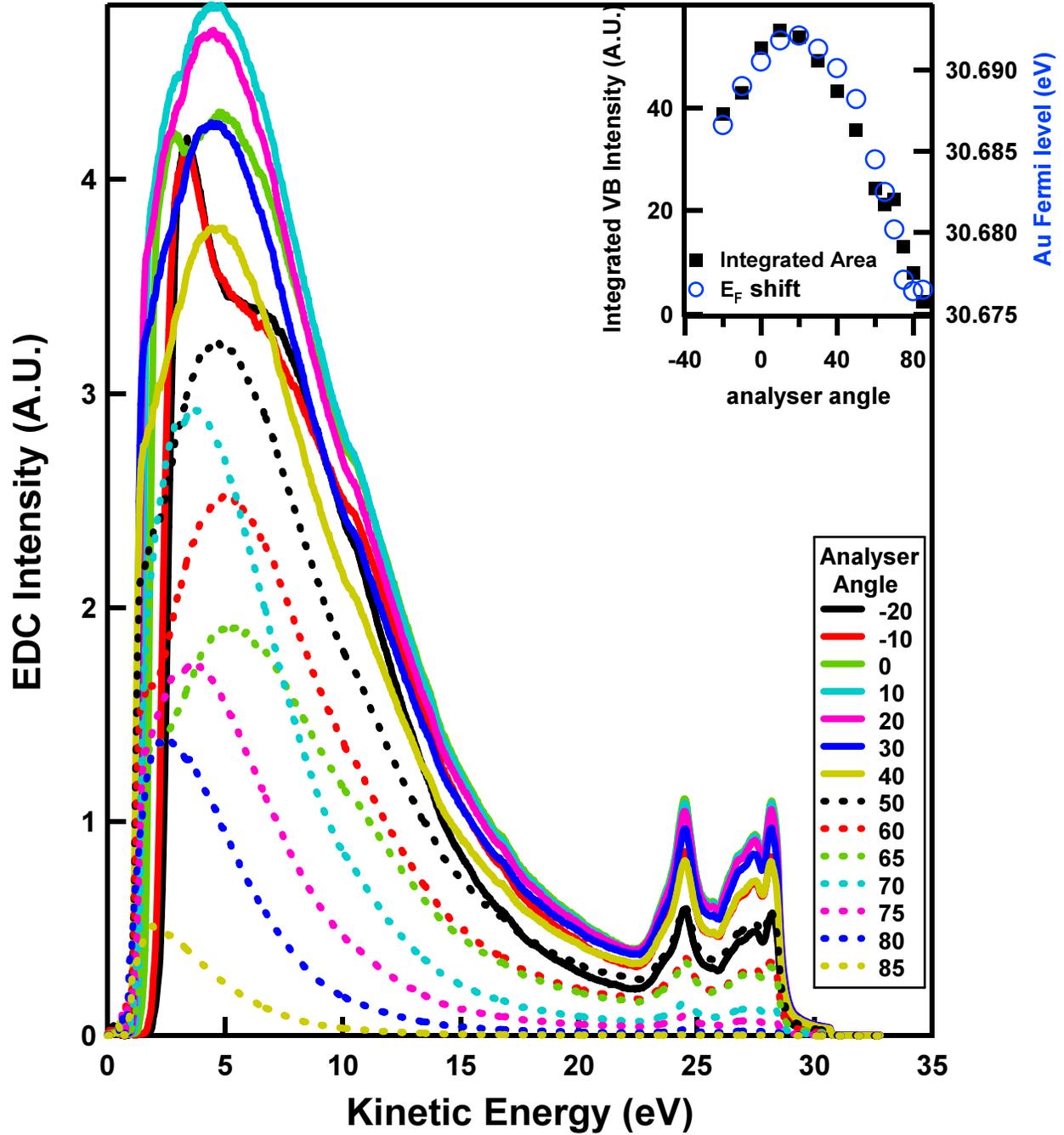}
\end{center}
\caption{Large energy-range valence band of Au measured at
different analyzer angle $\alpha$. The sample tilt angle is
$\phi$=37 degrees. In the inset shows the integrated spectral
weight over the entire energy range of 5$\sim$35 eV as a function
of the analyzer angle $\alpha$ (black solid square). For
comparison, the Fermi level as a function of the analyzer angle
measured under similar condition is also plotted (blue circle).
 }
\end{figure}

\begin{figure}[tbp]
\begin{center}
\includegraphics[width=0.8\columnwidth,angle=-90]{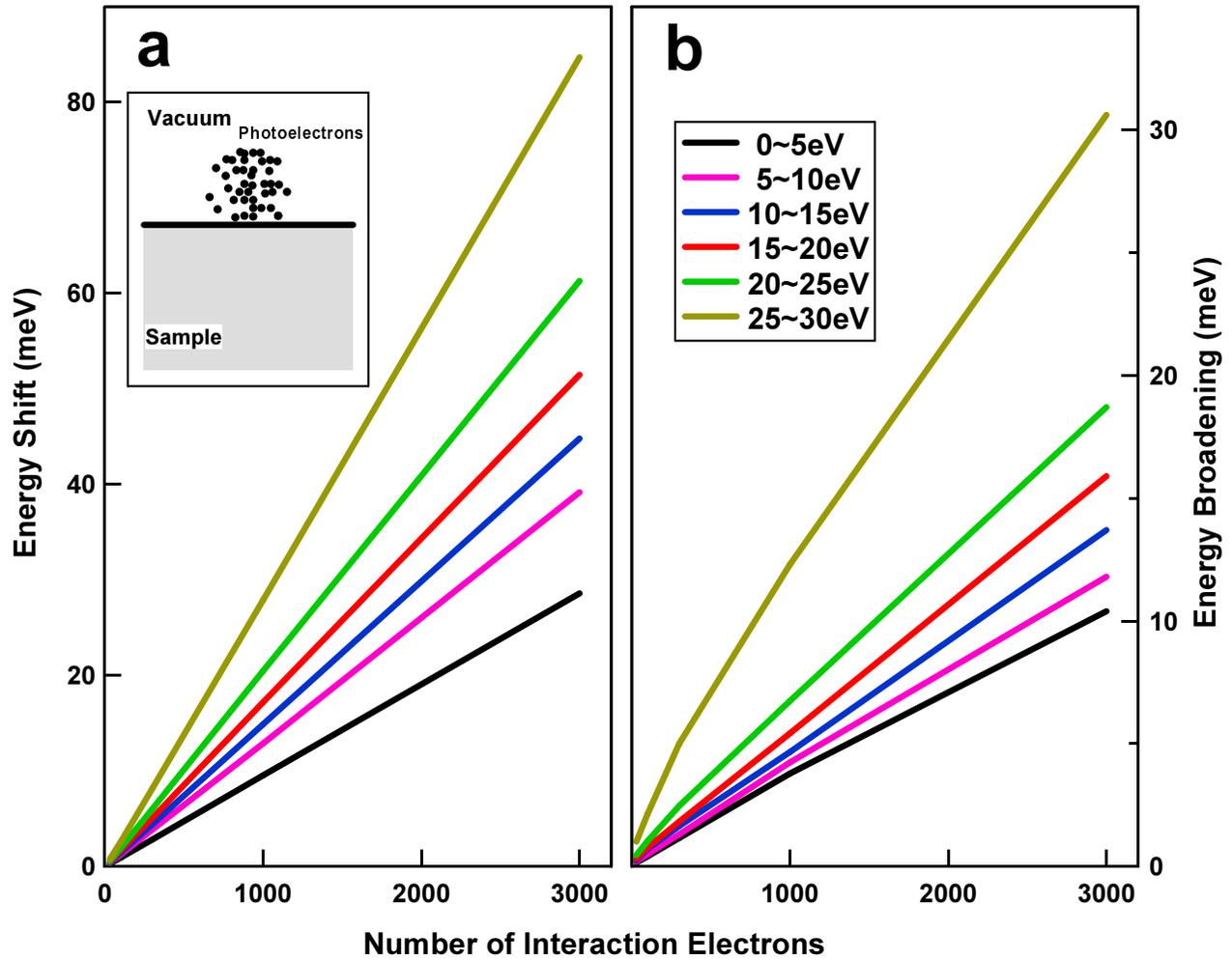}
\end{center}
\caption{Numerical calculations of space charge effect. The pulse
length is 60 ps and the spot size is 0.43mm$\times$0.42mm. (a).
The energy shift as a function of the number of interaction
electrons from the space charge effect. The test electron has an
energy of 30eV. Each curve represents the energy shift from the
interaction electrons with an energy range of 0$\sim$5 eV,
5$\sim$10 eV, 10$\sim$15 eV, 15$\sim$20 eV, 20$\sim$25 eV and
25$\sim$30eV. (b). The corresponding energy broadening from the
space charge effect for those six different energy regions. The
inset of (a) illustrates the space charge effect that shows a
number of electrons (solid circles) in a pulse escaping from the
sample surface. }
\end{figure}

\begin{figure}[tbp]
\begin{center}
\includegraphics[width=0.85\columnwidth]{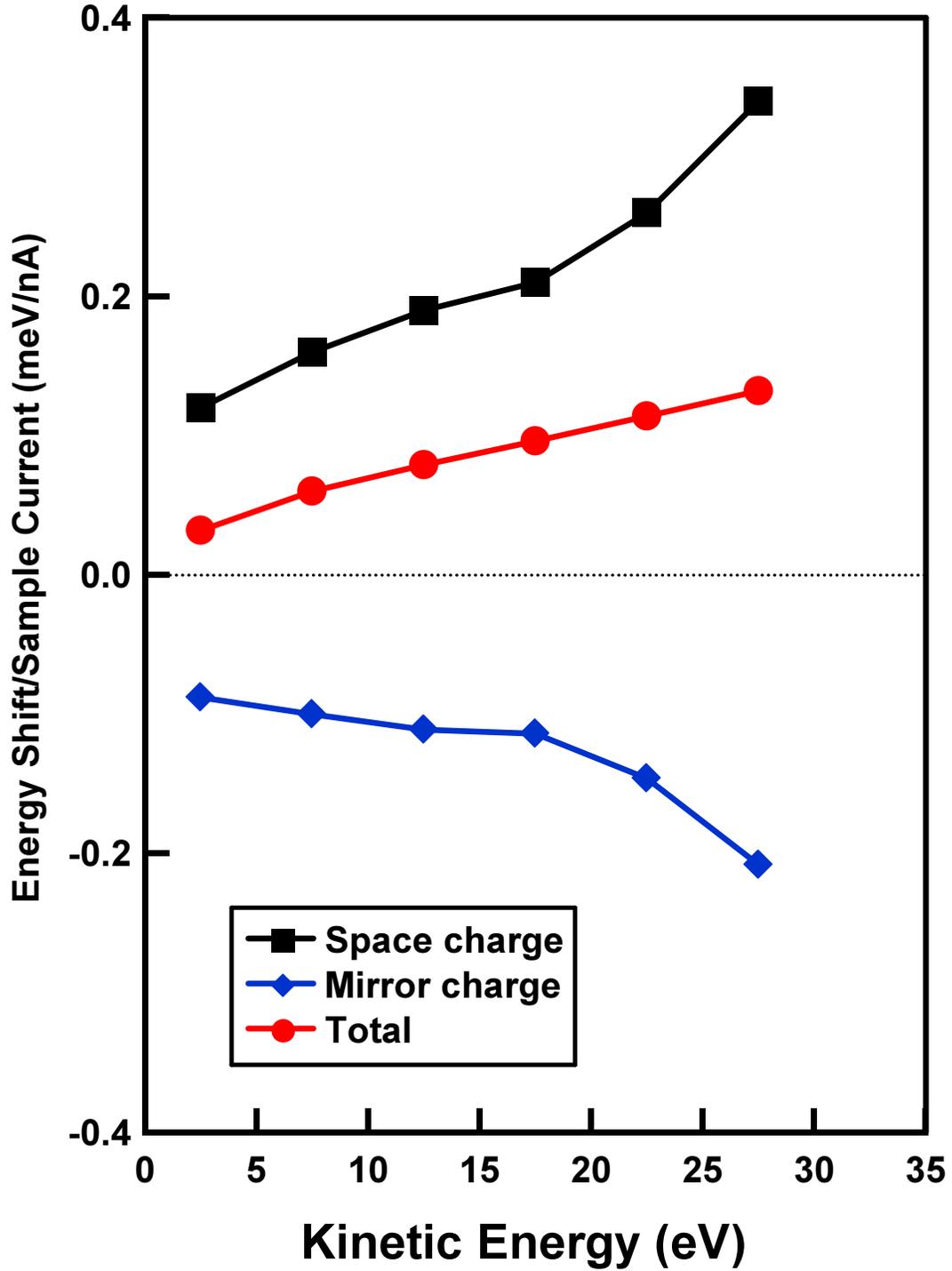}
\end{center}
\caption{The energy shift as a function of the interaction
electron energy distribution for the space charge only, the mirror
charge only, and both the space charge and the mirror charge.
These values are obtained by extracting the slope of curves in
Fig. 10a and 12a. }
\end{figure}

\begin{figure}[tbp]
\begin{center}
\includegraphics[width=0.8\columnwidth,angle=-90]{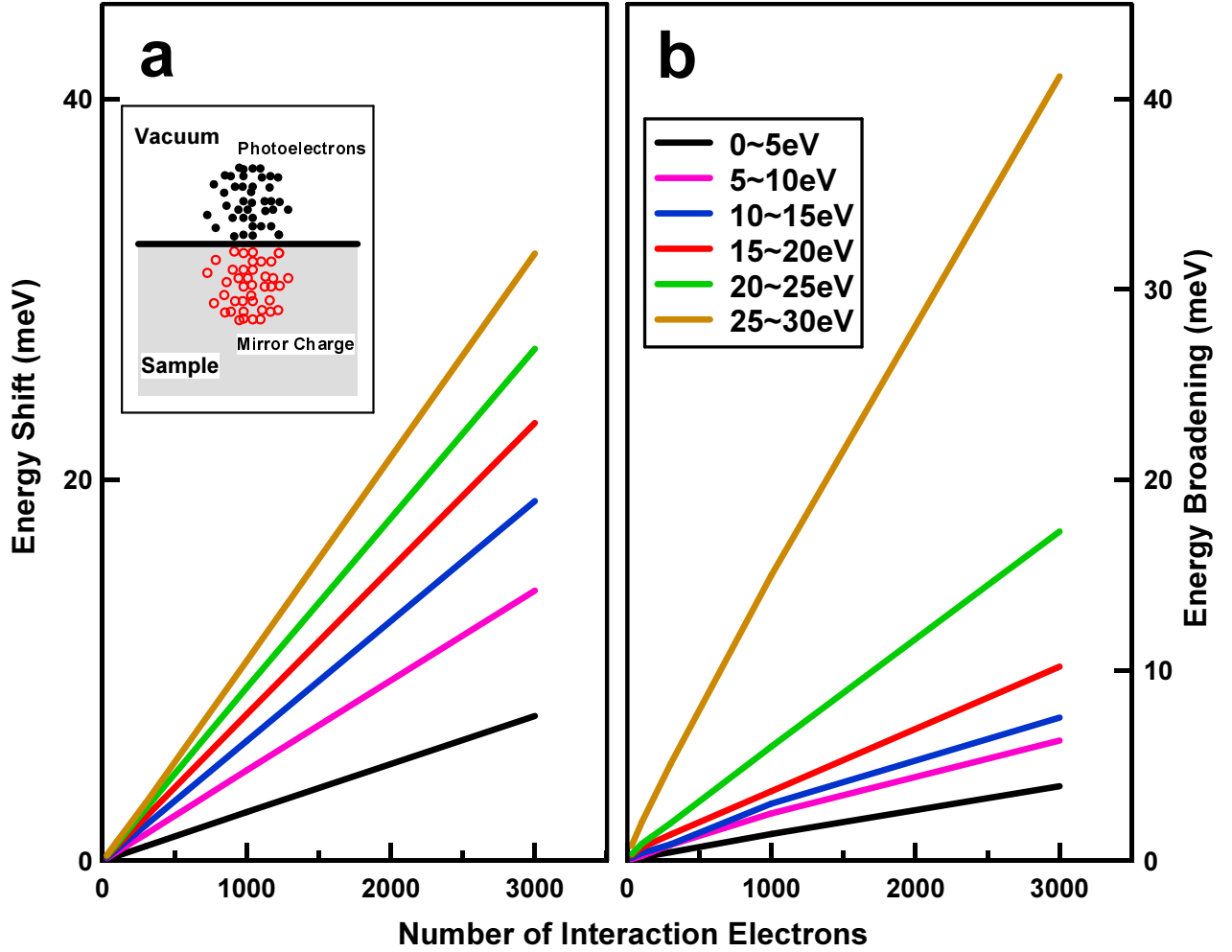}
\end{center}
\caption{Numerical calculations of combined space and mirror
charge effect. The pulse length is 60 ps and the spot size is
0.43mm$\times$0.42mm. (a). The energy shift as a function of the
number of the interaction electrons from both space charge and
mirror charge effects. The test electron has an energy of 30eV.
Each curve represents the energy shift from the interaction
electrons with an energy range of 0$\sim$5 eV, 5$\sim$10 eV,
10$\sim$15 eV, 15$\sim$20 eV, 20$\sim$25 eV and 25$\sim$30eV. (b).
The corresponding energy broadening from both space charge and
mirror charge for those six different energy regions. The inset of
(a) schematically shows the mirror charge effect: a pulse of
electrons (solid circles) escaping from the sample surface and
each electron has a mirror charge (open circles) inside the
sample. }
\end{figure}

\begin{figure}[tbp]
\begin{center}
\includegraphics[width=\columnwidth]{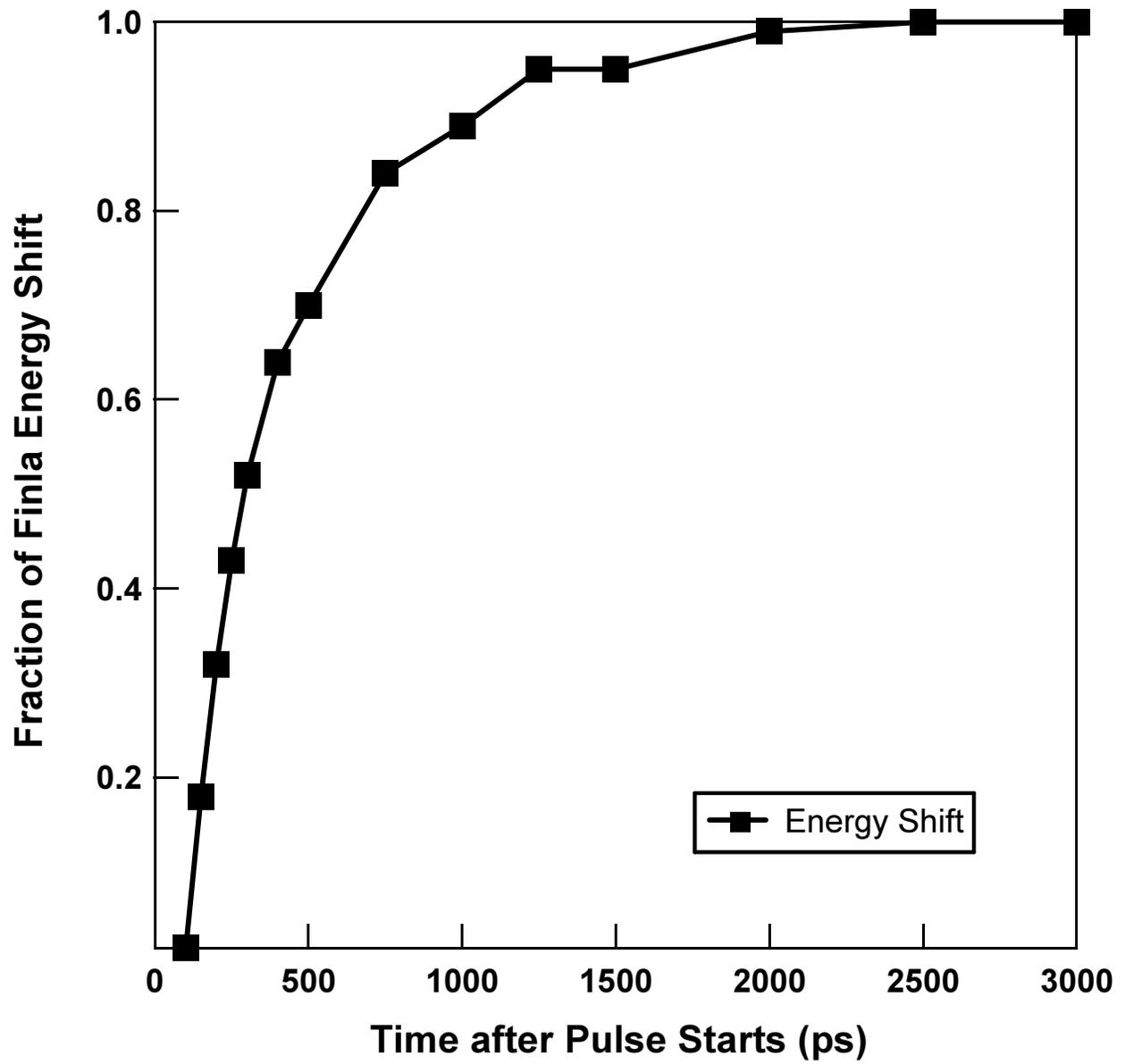}
\end{center}
\caption{Time evolution of the energy shift. The data are obtained
by averaging for 9, 30, 90 and 300 electrons/pulse. }
\end{figure}

\begin{figure}[tbp]
\begin{center}
\includegraphics[width=0.60\columnwidth]{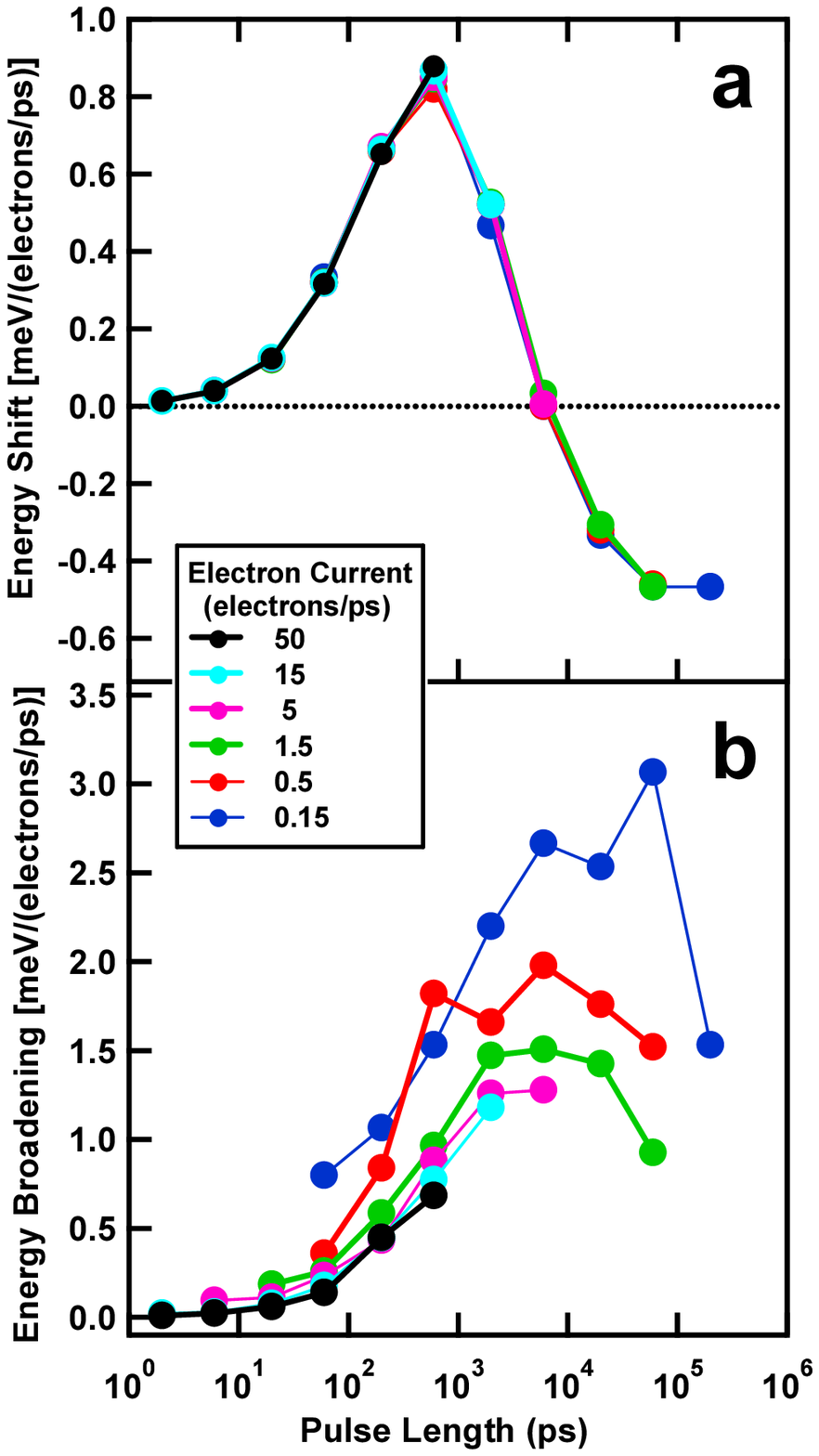}
\end{center}
\caption{(a)Energy shift and (b)broadening per electron current
(the unit is electrons per picosecond (e/ps))as a function of
pulse length at different electron currents. The spot size is
0.43mm$\times$0.42mm.  For the energy shift, all curves overlap
with each other, indicating that the energy shift is proportional
to the electron current. But for the energy broadening, they do
not strictly overlap with each other, particularly at longer pulse
length.}
\end{figure}


\begin{thebibliography}{99}

\bibitem{Huefner} S. Huefner, Photoemission Spectroscopy:
Principles and Applications (Springer-Verlag, Berlin, 1995).
\bibitem{SKeven} Angle-Resolved Photoemission: Theory and Current
Applications, edited by S. D. Kevan, (Elsevier, The Netherlans,
1992).
\bibitem{ScienceIssue} Special issue of Science {\bf
288}, No. 5465, (2000).
\bibitem{SpecialJESRP} Special issue of J.
Electron Spectroscopy and Related Phenomena, {\bf {117-118}}
1(2001).
\bibitem{Damascelli} A. Damascelli, Z. Hussain and Z.-X. Shen, Rev.
Modern Phys. {\bf 75}, 473(2003).
\bibitem{Chainani} A. Chainani et al., Phys. Rev. Lett. {\bf 85},
1966(2001).
\bibitem{Armitage} N. P. Armitage et al., Phys. Rev. Lett. {\bf 86},
1126(2001); T. Sato et al., Science {\bf 291}, 1517(2001).
\bibitem{Bjorn} B. Wannberg, P. Baltzer and S. Shin, preprint (2000).
\bibitem{Boersch} H. Boersch, Z. Physik {\bf 139},  115 (1954).
\bibitem{UHofer} U. Hofer et al., Science {\bf 277}, 1480 (1997);
P. M. Echenique and J. B. Pendry, Progress in Surf. Sci. {\bf 32},
111 (1989).
\bibitem{Thanks} We thank A. Fujimori, J. Bozek and S.
Sodergren for stimulating discussions.  The experiment was
performed at the ALS of LBNL, which is operated by the DOE's
Office of BES, Division of Material Science, with contract
DE-FG03-01ER45929-A001. The division also provided support for the
work at SSRL with contract DE-FG03-01ER45929-A001. The work at
Stanford was supported by NSF grant DMR-0304981 and ONR grant
N00014-98-1-0195-P0007, and the work at Colorado was supported by
NSF grant DMR 0402814  and DOE grant DE-FG02-03ER46066.


\end{thebibliography}
\end{document}